\documentclass[twocolumn,showkeys,aps,prb,showpacs]{revtex4-1}
\usepackage{graphicx}
\usepackage[CJKbookmarks,dvipdfm,colorlinks,linkcolor=blue,citecolor=blue]{hyperref}

\begin{document}

\title{Biaxial strain tuned electronic structures and power factor in Janus Transition Metal Dichalchogenide monolayers}

\author{San-Dong Guo}
\affiliation{Department of Physics, School of Sciences, China University of Mining and
Technology, Xuzhou 221116, Jiangsu, China}
\begin{abstract}
 Tuning physical properties of transition metal dichalcogenide (TMD) monolayers  by strain engineering have most widely studied, and recently
 Janus TMD monolayer MoSSe  has been synthesized.
In this work, we systematically study biaxial strain dependence of electronic structures and  transport properties of Janus TMD  MXY (M = Mo or W, X/Y = S, Se, or Te) monolayer by using generalized gradient approximation (GGA) plus spin-orbit coupling (SOC). It is found that SOC has a noteworthy detrimental influence on  power factor  in  p-type MoSSe, WSSe, n-type WSTe, p-type MoSeTe and WSeTe, and has a negligible influence on one in n-type MoSSe, MoSTe,  p-type WSTe and n-type MoSeTe. These can be understood by  considering SOC  effects on their  valence and conduction bands. For all six monolayers, the energy band gap firstly increases, and then decreases, when strain changes from compressive one to tensile one. It is found that strain can tune strength of bands convergence of both valence and conduction  bands by changing the numbers and relative position of valence band extrema (VBE) or conduction band extrema (CBE), which can produce   very important effects on their electronic transport properties.
By applying appropriate compressive or tensile strain, both  n- or p-type Seebeck coefficient   can be enhanced  by  strain-induced  band convergence, and then the power factor can be improved. Our works further enrich studies on  strain dependence of electronic structures and  transport properties of new-style TMD monolayers, and motivate farther experimental works.

\end{abstract}

\pacs{72.15.Jf, 71.20.-b, 71.70.Ej, 79.10.-n ~~~~~~~~~~~~~~~~~~~~~~~~~~~~~~~~~~~Email:sandongyuwang@163.com}

\maketitle

\section{Introduction}
Due to direct hot-electricity conversion without moving parts, thermoelectric materials have enormous potential to solve energy issues, and the efficiency of thermoelectric conversion can be measured by the dimensionless  figure of merit\cite{s1,s2}, $ZT=S^2\sigma T/(\kappa_e+\kappa_L)$,  in which  S is the Seebeck coefficient,   $\sigma$ is electrical conductivity,  T is absolute  temperature,  $\kappa_e$ and $\kappa_L$ are the electronic and lattice thermal conductivities, respectively. Based on the expression of $ZT$, an excellent efficiency of thermoelectric conversion requires high power factor ($S^2\sigma$) and low thermal conductivity ($\kappa=\kappa_e+\kappa_L$). However,  the  S and  $\sigma$ are oppositely  proportional to the carrier concentration.
Due to simultaneously increasing $S^2\sigma$ and decreasing $\kappa$, low-dimensional materials may have potential advantages
 in improving $ZT$\cite{q1,q2,q3}.

Since the discovery of graphene\cite{q6}, two-dimensional (2D) materials have been
attracting increasing attention, such as TMD, group-VA, group IV-VI and group-IV  monolayers\cite{q7,q8,q9,q10,q11}.
The  heat transport properties of these 2D materials  have been widely studied, such as  TMD,  orthorhombic group IV-VI, group-VA, $\mathrm{SnSe_2}$,  ATeI (A=Sb or Bi) and $\mathrm{TiS_2}$  monolayers  \cite{q12,q13,q14,q15,q16,q17,q18,q19}.  In semiconducting TMD monolayers $\mathrm{MX_2}$ (M=Zr, Hf,  Mo, W or Pt; X=S, Se, or Te), the SOC is proved to be very important for electronic transport properties\cite{q20}.  Strain effects on the electronic structures and heat transport properties of TMD monolayers have been widely investigated both in theory and experiment.  A semiconductor-to-metal transition can be observed by a small compressive strain (about 3\%) in $\mathrm{PtTe_2}$, compared with $\mathrm{MoS_2}$ with very large strain\cite{q21,q21-1}.
For $\mathrm{MoS_2}$, the  significantly enhanced power factor can be observed in  n(p)-type doping by compressive (tensile) strain at the critical strain of
 direct-indirect gap transition\cite{q21-2}.
It is found that  tensile strain can improve thermoelectric properties of $\mathrm{ZrS_2}$, $\mathrm{PtSe_2}$ and  $\mathrm{PtTe_2}$ by enhancing $S^2\sigma$ and reducing $\kappa_L$\cite{q21,q22,q23}.
\begin{figure}
  \includegraphics[width=7.0cm]{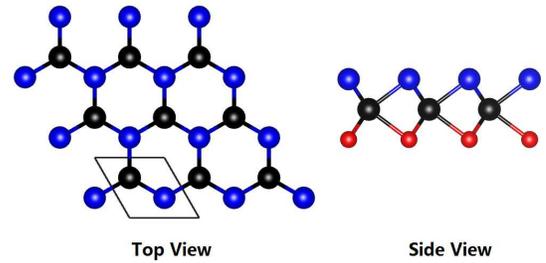}
  \caption{(Color online) The schematic crystal structure of Janus MXY (M = Mo or W, X/Y = S, Se, or Te) monolayer. The black balls represent M atoms, and the red and blue balls for X/Y atoms.}\label{t0}
\end{figure}
\begin{figure}
  \includegraphics[width=8cm]{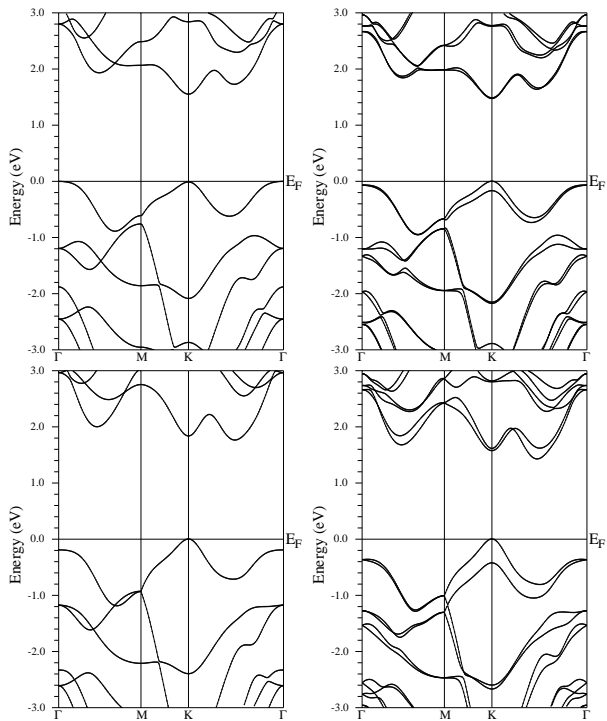}
\caption{The energy band structures  of MoSSe (Top) and WSSe (Bottom)  using GGA (Left) and GGA+SOC (Right). }\label{t1}
\end{figure}

Recently,  Janus monolayer MoSSe has been  experimentally achieved  by  breaking the out-of-plane structural
symmetry of $\mathrm{MoS_2}$,  replacing the top S atomic layer  with Se atoms\cite{p1}.
It is found that Janus MoSSe monolayer can be used as a potential wide solar-spectrum water-splitting photocatalyst with a low carrier recombination rate\cite{p1-1}.  In monolayer and multilayer Janus TMD  MXY (M = Mo or W, X/Y = S, Se, or Te),
the  strong piezoelectric effects have been observed
  by first-principles calculations\cite{p2}. It is found that the carrier mobility in monolayer MoSSe is relatively low,  but the bilayer or trilayer structures show a quite high electron/hole carrier mobility\cite{p2-1}.
 Electronic and optical properties have been investigated  in pristine Janus MoSSe and WSSe monolayers,   as well as their  vertical and lateral heterostructures\cite{p3}. It is found  that the $\kappa_L$ of MoSSe monolayer is higher than that of  $\mathrm{MoSe_2}$ monolayer, but is  very lower than that of $\mathrm{MoS_2}$ monolayer\cite{p4}.  Calculated results show that ZrSSe monolayer predicted with  the 1T phase has better n-type thermoelectric properties than monolayer  $\mathrm{ZrS_2}$\cite{p4-1}.

In this work, the biaxial strain dependence of electronic structures and  transport properties of  Janus TMD  MXY (M = Mo or W, X/Y = S, Se, or Te) monolayers  are studied by  first-principles calculations and   Boltzmann equation. It is very crucial for  Janus TMD monolayers to include SOC for attaining reliable electronic structures and  transport properties, which is similar with TMD monolayers\cite{q20,q21,q21-2}.
 For all six  Janus TMD monolayers,  the energy band gap  shows a nonmonotonic up-and-down behavior with increasing strain, while the spin-orbit splitting at $K$ point monotonically increases.
 Calculated results show  that strain can tune strength of bands convergence of  valence (conduction) bands by changing the numbers and relative position of VBE  (CBE), which can obviously affect their electronic transport properties. Both  n- or p-type Seebeck coefficient   can be enhanced  by
 applying appropriate compressive or tensile strain, and then the power factor can be improved. Similar strain-improved power factor can also be found in TMD monolayers\cite{q21,q21-2,q22,q23}.

\begin{table}[!htb]
\centering \caption{For MXY (M = Mo or W, X/Y = S, Se, or Te) monolayer, the  lattice constants\cite{p2} $a$ ($\mathrm{{\AA}}$); the calculated energy band gaps  using GGA $G$ (eV) and GGA+SOC $G_{so}$ (eV); $G$-$G_{so}$ (eV);   the spin-orbit splitting value $\Delta$ (eV) at K point in the valence bands around the Fermi level. }\label{tab}
  \begin{tabular*}{0.48\textwidth}{@{\extracolsep{\fill}}ccccccc}
  \hline\hline
Name& $a$ &  $G$& $G_{so}$&$G$-$G_{so}$& $\Delta$\\\hline\hline
MoSSe&3.252&1.55&1.47&0.08& 0.168\\\hline
MoSTe&3.327&1.17&1.14&0.03& 0.181\\\hline
MoSeTe&3.394&1.34&1.22&0.12&0.196\\\hline
WSSe&3.220&1.76&1.43&0.33& 0.426\\\hline
WSTe&3.325&1.35&1.21&0.14& 0.396\\\hline
WSeTe&3.391&1.67&1.08&0.59&0.433\\\hline\hline
\end{tabular*}
\end{table}

The rest of the paper is organized as follows. In the next section, we shall
describe computational details about electronic structures and transport properties. In the third section, we shall present strain dependence of the electronic structures and  transports  properties of  Janus TMD  MXY (M = Mo or W, X/Y = S, Se, or Te) monolayers. Finally, we shall give our discussions and conclusion in the fourth
section.

\section{Computational detail}
A full-potential linearized augmented-plane-waves method
within the density functional theory (DFT) \cite{1}  is used to investigate strain dependence of
electronic structures  of  MXY (M = Mo or W, X/Y = S, Se, or Te) monolayer, as implemented in
the WIEN2k  package\cite{2}.
We employ the popular GGA of Perdew, Burke and  Ernzerhof  (GGA-PBE)\cite{pbe} for the
exchange-correlation potential, and the  internal position parameters are optimized with a force standard of 2 mRy/a.u..
The SOC was included self-consistently \cite{10,11,12,so}, which can produce  important effects  on both electronic structure and transport coefficients.
 To attain reliable results, we use 5000 k-points in the
first Brillouin zone (BZ) for the self-consistent calculation,  make harmonic expansion up to $\mathrm{l_{max} =10}$ in each of the atomic spheres, and set $\mathrm{R_{mt}*k_{max} = 8}$.
The self-consistent calculations are
considered to be converged when the integration of the absolute
charge-density difference between the input and output electron
density is less than $0.0001|e|$ per formula unit, where $e$ is
the electron charge.

Based on calculated energy band
structures, transport coefficients, such as  Seebeck coefficient and electrical conductivity, are performed through solving Boltzmann
transport equations within the constant
scattering time approximation (CSTA) as implemented in
BoltzTrap\cite{b}. To achieve the convergence results, the parameter LPFAC is set to 40.
The accurate transport coefficients need dense k-point meshes, and at least 2400 k-points is used in the  irreducible BZ for the energy band calculation.
It is noted that, for 2D material, the calculated  electrical conductivity  depends on the length of unit cell along z direction\cite{2dl}.  They  should be normalized by multiplying $Lz/d$, in which  $Lz$ is the length of unit cell along z direction,  and $d$ is the thickness of 2D material. It is well known that the $d$  is not well defined like graphene.  In this work, the $Lz$=20 $\mathrm{{\AA}}$  is used as $d$.

\begin{figure}
  \includegraphics[width=7.0cm]{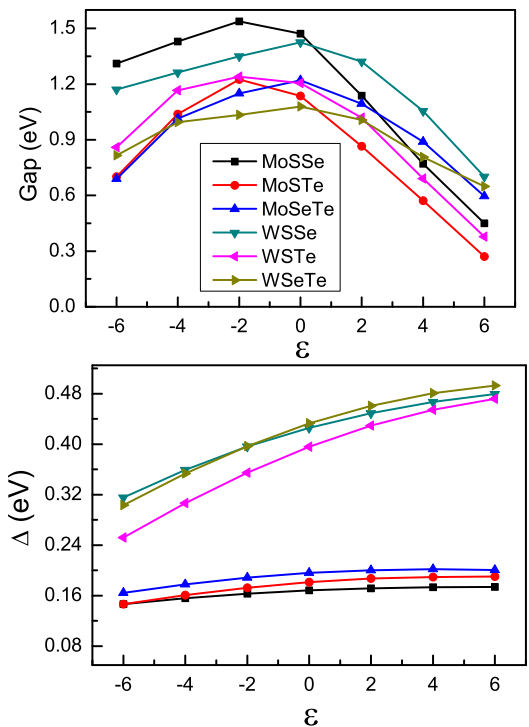}
  \caption{(Color online)For MXY (M = Mo or W, X/Y = S, Se, or Te) monolayer, the energy band gap (Gap) and  spin-orbit splitting value ($\Delta$)  at high symmetry K point  as a function of  $\varepsilon$ by using GGA+SOC.}\label{t2}
\end{figure}

\section{MAIN CALCULATED RESULTS AND ANALYSIS}
The structure of Janus MXY (M = Mo or W, X/Y = S, Se, or Te) monolayer (\autoref{t0}) is similar to  $\mathrm{MX_2}$  monolayer with the 2H phase,
which contains three atomic sublayers with M layer sandwiched between X and Y layers.  Compared with $\mathrm{MX_2}$, the Janus MXY
monolayer lacks the reflection symmetry with respect to the central metal M atoms.
With the sandwiched S-Mo-Se structure, Janus TMD monolayer MoSSe  has been experimentally achieved by replacing the top S atomic layer in  $\mathrm{MoS_2}$ with Se atoms\cite{p1}. To avoid spurious interaction between neighboring layers, the unit cell  of  Janus MXY monolayer,  containing  one M, one X and one Y atoms, is constructed with the vacuum region of more than 18 $\mathrm{{\AA}}$. The optimized lattice constants\cite{p2} for MXY are listed in \autoref{tab} using GGA.

\begin{figure*}
  \includegraphics[width=15cm]{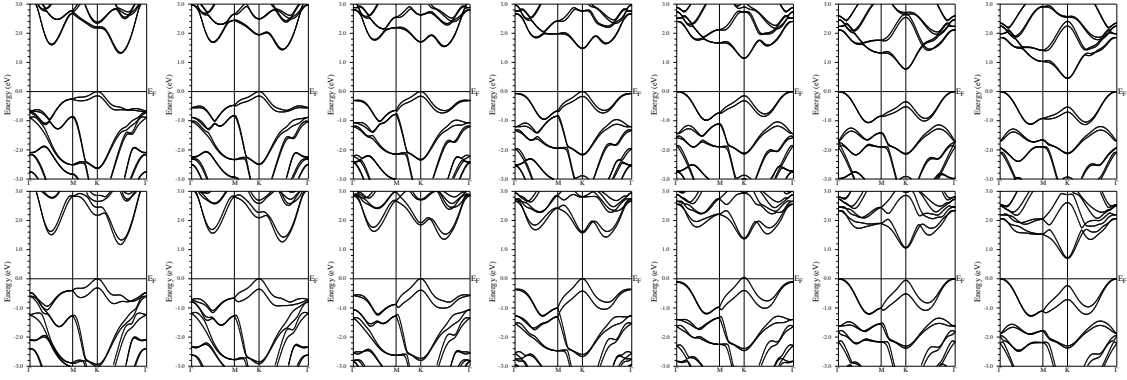}
\caption{The energy band structures  of MoSSe (Top) and WSSe (Bottom)  with $\varepsilon$ changing from -6\% to 6\%  using GGA+SOC, and the strain increment for 2\%. }\label{t3}
\end{figure*}

\begin{figure*}
  \includegraphics[width=15cm]{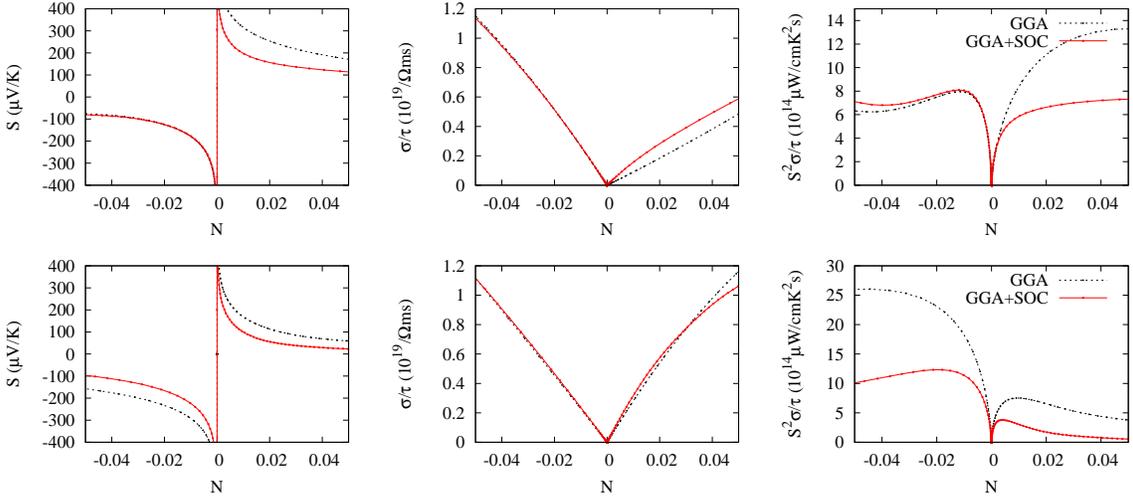}
  \caption{(Color online) the room-temperature transport coefficients of MoSSe (Top) and WSSe (Bottom)  as a function of doping level (N) using GGA and GGA+SOC: Seebeck coefficient S, electrical conductivity with respect to scattering time  $\mathrm{\sigma/\tau}$  and   power factor with respect to scattering time $\mathrm{S^2\sigma/\tau}$.}\label{t4}
\end{figure*}

  It has been proved that SOC can produce important effects on electronic structures for  $\mathrm{MX_2}$ (M=Zr, Hf,  Mo, W or Pt; X=S, Se, or Te), and further influences their thermoelectric properties\cite{q20,q21,q21-2,q23}. Due to similar crystal structure and element type between TMD and  Janus TMD monolayers, the SOC is included for all calculations of Janus TMD monolayers.  \autoref{t1} shows the calculated energy bands for monolayer MoSSe and WSSe with GGA and GGA+SOC, and   FIG.1 and FIG.2 in the Supporting Information (SI) show ones of monolayer MoSTe, WSTe, MoSeTe and WSeTe. For monolayer MoSSe, the indirect gap of 1.55 eV is calculated with valence band maximum (VBM) at $\Gamma$ point and conduction band minimum (CBM) at K point using GGA. A second maxima appears  at K point, which is  0.01 eV lower than  VBM.
When the SOC is considered, the VBM changes from $\Gamma$ point to K point with a direct gap of 1.47 eV, and the energy difference between $\Gamma$ and K is  0.07 eV. It is noted that these results  sensitively depend on lattice constants.
For WSSe, the CBM is along the $\Gamma$-K direction, and an indirect gap of 1.76 eV (1.43 eV) using GGA (GGA+SOC) is defined with the
VBM at the $\Gamma$ point. For MoSTe and WSTe, the CBM and VBM are
located along the $\Gamma$-K direction and  at  $\Gamma$ point. The MoSeTe and WSeTe have indirect gaps with the CBM and VBM  along the $\Gamma$-K direction and  at  K point. The GGA gaps, GGA+SOC gaps and the differences between them are shown in \autoref{tab}. It is found that the gaps with GGA+SOC are smaller than ones with GGA for all materials, which is caused by spin-orbit splitting. It is found that the  Rashba spin-orbit splitting exists at $\Gamma$ point of valence bands because of lacking the inversion symmetry. The gap difference between GGA and GGA+SOC can reflect the SOC influences on the conduction bands, and
the larger gap decrease means the stronger SOC.  The SOC effects on the valence bands near Fermi level can be described by  spin-orbit splitting at  the K point, which  are summarized  in \autoref{tab}. It is clearly seen that WXY has larger spin-orbit splitting than MoXY.

Both in theory and in experiment, strain effects on the energy band structures and  transport properties of TMD monolayers have been widely investigated\cite{q21,q21-2,q22,q23,qin1}.
Here,  biaxial strain effects on the electronic structures and  electronic transport coefficients of   MXY (M = Mo or W, X/Y = S, Se, or Te) monolayer are studied. To simulate biaxial strain, $\varepsilon=(a-a_0)/a_0$ is defined, where $a_0$ is the unstrained lattice constant. $\varepsilon$$<$0 means  compressive strain, while  $\varepsilon$$>$0 implies tensile strain.
Using GGA+SOC, the energy band gap  and   spin-orbit splitting value at  K point  in the valence bands around the Fermi level  as a function of  $\varepsilon$   are plotted in \autoref{t2}. For all materials, the energy band gap  firstly increases, and then decreases, when $\varepsilon$ changes from -6\% to 6\%.
Similar strain dependence of energy band gap can also be found in TMD monolayers\cite{q21,q21-2,q22,q23}. As $\varepsilon$ increases, the spin-orbit splitting at K point monotonically increases, and the change is 0.027$-$0.044 eV for MoXY and 0.164$-$0.220 eV for WXY, which means the spin-orbit splitting has stronger dependence on strain for WXY than MoXY. With increasing strain, the trend of spin-orbit splitting  is consistent with one of  $\mathrm{MoS_2}$\cite{q21-2}, but is opposite to one of  $\mathrm{PtSe_2}$\cite{q23} or  $\mathrm{PtTe_2}$\cite{q21}.

For monolayer MoSSe and WSSe,  the related energy band structures with strain from -6\% to 6\% are also shown  in \autoref{t3} using GGA+SOC,
and   FIG.3 and FIG.4 in the SI show ones of monolayer MoSTe, WSTe, MoSeTe and WSeTe. For all materials, there are  some  VBE and CBE around the Fermi level. It is found that strain can tune the numbers and relative position of VBE or CBE, which can produce   very important influences on their electronic transport properties. The compressive strain can reduce the numbers of CBE from three to two, and tensile strain from three to one. Both compressive and tensile strain can change relative position of VBE. In a word, strain can tune strength of bands convergence of both conduction and valence bands. The similar phenomenon can also be observed in TMD monolayers\cite{q21,q21-2,q22,q23,qin1}.

\begin{figure*}
    \includegraphics[width=15cm]{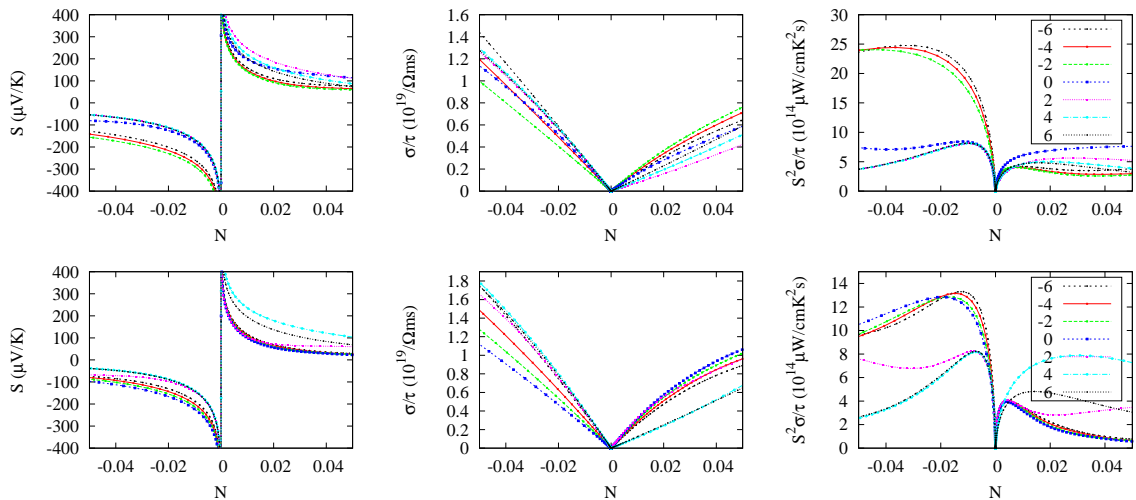}
  \caption{(Color online) the room-temperature transport coefficients of MoSSe (Top) and WSSe (Bottom)  as a function of doping level (N) using GGA+SOC  with $\varepsilon$ changing from -6 to 6: Seebeck coefficient S, electrical conductivity with respect to scattering time  $\mathrm{\sigma/\tau}$  and   power factor with respect to scattering time $\mathrm{S^2\sigma/\tau}$. }\label{t5}
\end{figure*}

 The transport coefficients calculations are performed, based on CSTA Boltzmann theory within rigid band approach.  The calculated electrical conductivity $\mathrm{\sigma/\tau}$ depends  on scattering time, while  Seebeck coefficient S is independent of scattering time.
 By simply moving  the position of Fermi level, the doping effects can be  simulated.  When  the Fermi level is shifted  into conduction  (valence) bands, the n(p)-type doping is achieved with negative (positive) doping levels, giving the negative (positive) Seebeck coefficient. For monolayer MoSSe and WSSe, at room temperature, the  Seebeck coefficient S,  electrical conductivity with respect to scattering time  $\mathrm{\sigma/\tau}$ and  power factor with respect to scattering time $\mathrm{S^2\sigma/\tau}$  as  a function of doping level (N)   using GGA and GGA+SOC are plotted  in \autoref{t4}, and FIG.5 and FIG.6 in the SI show ones of monolayer MoSTe, WSTe, MoSeTe and WSeTe.
For MoXY,  a detrimental influence on p-type  S  can be induced by SOC,  while  a neglectful effect on S (absolute value) in n-type doping can be observed.
The SOC  can lift the valence band degeneracy near the K point,  which reduces  slope of density of states (DOS) of valence bands near the energy gap, giving rise to reduced Seebeck coefficient. However, the weak SOC effects on conduction bands near the Fermi level are observed, leading to  a neglectful effect on n-type S.
For WXY,  a reduced  influence on both n- and  p-type  S can be observed at the presence of SOC, which can be explained by SOC-induced spin-orbit splitting of both conduction and valence bands, reducing slope of DOS  near the energy gap.
In n-type doping, the power factor of WXY with GGA+SOC is smaller than one using GGA. For Mo/WSSe and Mo/WSeTe, p-type power with GGA+SOC is lower than one with GGA. These can be understood by SOC effects on S and $\mathrm{\sigma/\tau}$. It is noted that theses results depend on the lattice constants. When the SOC is considered, the strength of bands convergence is enhanced, and the S would be improved, producing enhanced power factor, which has been observed in TMD monolayer $\mathrm{WX_2}$  (X=S, Se and Te)\cite{q20}.

At 300 K, the biaxial strain dependence of S,  $\mathrm{\sigma/\tau}$ and $\mathrm{S^2\sigma/\tau}$ of  monolayer MoSSe and WSSe    are shown  in \autoref{t5} using GGA+SOC, and  FIG.7 and FIG.8 in the SI show ones of monolayer MoSTe, WSTe, MoSeTe and WSeTe.
The complex strain  dependence  of  transport coefficients are observed, which is because  their  energy band structures are  sensitively dependent on strain.
Strain-enhanced S can be understood  by strain-driven  accidental degeneracies, namely bands convergence.  For example MoSSe, in considered  n-type doping range, the largest S  can be observed  with -2\% strain  due to the near degeneracy  among  CBE along   K-$\Gamma$,   along  $\Gamma$-M and  at K point.  In fact, the MoSSe with -6\% and -4\% strain have similar S with one with -2\% due to the bands convergence of CBE along   K-$\Gamma$ and   $\Gamma$-M, leading to very large n-type power factor. In p-type doping of WSSe,  S  reaches the largest values  with 4\% strain due to  the energy levels of K  and $\Gamma$ points being  more close, leading to largest p-type power factor.  It is found that the  $\mathrm{\sigma/\tau}$ and S show usually opposite strain dependence.
For MoSSe, WSSe, MoSTe and MoSeTe, strain-enhanced n-type  power factor is larger than p-type  one, while it is opposite for WSTe and WSeTe.
An upper limit of $ZT$, neglecting $\kappa_L$,  can be defined as $ZT_e=S^2\sigma T/\kappa_e$.
The   $\mathrm{\kappa_e}$ relates to   $\mathrm{\sigma}$ via the Wiedemann-Franz law: $\kappa_e=L\sigma T$, and then $ZT_e=S^2/L$, where L is the Lorenz number.
Therefore, the power factor is improved by enhanced S induced by strain, which is  beneficial to better thermoelectric properties.
Strain-improved power factor can also be observed in TMD monolayers\cite{q21,q21-2,q22,q23,qin1}.

\section{Discussions and Conclusion}
For TMD monolayers, the SOC  produces  a remarkable influence
on S caused by SOC-removed the band degeneracy, and further affects the power factor\cite{q20,q21,q21-2,q23}.
The SOC not only  can reduce the power factor, but can also obviously improve  one like  $\mathrm{WX_2}$  (X=S, Se and Te)\cite{q20}.
However, for unstrained  MXY (M = Mo or W, X/Y = S, Se, or Te) monolayer, only obviously reduced effect can be observed by SOC.
The strain-improved S can also found in TMD monolayers, such as  $\mathrm{MoS_2}$, $\mathrm{PtSe_2}$, $\mathrm{PtTe_2}$, $\mathrm{ZrS_2}$ and $\mathrm{ZrSe_2}$ \cite{q21,q21-2,q22,q23,qin1}, and the related
mechanism is similar with that of strain-enhanced  S of Janus TMD monolayers. Besides strain, electric field can also effectively tune the electronic structures
 of 2D materials, so it is possible to tune S of  Janus TMD monolayers by  electric field. The Janus TMD monolayers may have better thermoelectric properties than TMD monolayers due lower $\kappa_L$. It has been proved that the MoSSe (ZrSSe) has lower  $\kappa_L$ than  $\mathrm{MoS_2}$ ($\mathrm{ZrS_2}$)\cite{p4,p4-1}, and the ZrSSe has  enhanced n-type thermoelectric properties compared with monolayer  $\mathrm{ZrS_2}$\cite{p4-1}.

 In summary,  we  systematically study strain dependence of  electronic structures and  transport coefficients   of Janus MXY (M = Mo or W, X/Y = S, Se, or Te) monolayer, based mainly on the reliable first-principle calculations. Calculated results show that the inclusion of SOC is key for
energy band structures of Janus TMD monolayers, which has important effects on their  electronic   transport coefficients.
It is found that both compressive and tensile strain can tune the strength of bands
convergence by changing the numbers and relative position of VBE or CBE, producing important effects on their  electronic   transport coefficients.  For all Janus TMD monolayers, the S can be enhanced by choosing the appropriate compressive or tensile strain, and then the power factor can be improved.
 Our works will motivate farther  experimental studies, and studies of electronic transports of other Janus TMD monolayers.

\begin{acknowledgments}
This work is supported by the National Natural Science Foundation of China (Grant No. 11404391). We are grateful to the Advanced Analysis and Computation Center of CUMT for the award of CPU hours to accomplish this work.
\end{acknowledgments}

\end{document}